\documentclass{book}    

\usepackage{siunitx}
\usepackage{piers}  
\begin{document}

\title{Dog Bone Triplet Metamaterial Half Wave Plate}
\maketitle

\begin{authors}

{\bf I. Mohamed}, {\bf G. Pisano}, {\bf M. W. Ng}, {\bf V. Haynes}, {\bf and B. Maffei}\\
\medskip
Jodrell Bank Centre for Astrophysics, School of Physics and Astronomy\\The University of
Manchester, United Kingdom

\end{authors}

\begin{paper}

\begin{piersabstract}
Metamaterials are artificially made sub-wavelength structures arranged in periodic arrays. 
They can be designed to interact with electromagnetic radiation in many different and 
interesting ways such as allowing radiation to experience a negative refractive index (NRI). 
We have used this technique to design and build a quasi-optical Half Wave Plate (HWP) 
that exhibits a large birefringence by virtue of having a positive refractive index in one axis 
and a NRI in the other. Previous implementations of such NRI-HWP have been narrow band 
(\SIrange{\sim1}{3}{\percent}) due to the inherent reliance on needing a resonance to create 
the NRI region. We manufacture a W-band prototype of a novel HWP that uses the 
Pancharatnam method to extend the bandwidth (up to more than twice) of a usual 
NRI-HWP. Our simulated and experimentally obtained results despite their differences show 
that a broadening of a flat region of the phase difference is possible even with the initially 
steep gradient for a single plate.
\end{piersabstract}

\psection{Introduction}
Electromagnetic metamaterials are artificially created sub-wavelength structures and are 
known for their use in creating a negative refractive index (NRI), an effect first 
demonstrated in 2000 \cite{Smith2000}.

Wave plates are used to alter the polarisation of radiation passing through them. Rotating HWPs 
are used to rotate linear polarisations at twice their mechanical rotation speed. Conventionally 
made from birefringent materials such as sapphire or quartz, wave plates can also be constructed 
out of metal mesh grids \cite{Pisano2008}. Such constructions are advantageous due to the costs
and limited dimensions of the available birefringent materials at these frequencies and the ability 
to scale the designs to work at other frequencies. Their use is becoming increasingly important in 
Cosmic Microwave Background Radiation experiments where the detection of the weak B-mode 
polarisation is the current goal in observational cosmology.

The phase difference, $\Delta\phi$, between the two polarisations of radiation of frequency, $f$, 
after passing through a wave plate of thickness, $d$, is given by
\begin{equation}
\Delta\phi=\frac{2\pi d f}{c_0}\cdot\Delta n
\label{eq:dp}
\end{equation}
where $c_0$ is the speed of light in a vacuum and $\Delta n$ is the difference in the refractive 
indices of the birefringent materials two orthogonal axes. By designing a metamaterial with a 
geometry that has a NRI along one axis and a positive index along the other we are able to 
create a wave plate with very high birefringence: this allows the creation of relatively thin 
wave plates. Previous implementations \cite{Imhof2007,Weis2009} of metamaterial wave 
plates that have used NRI were narrow band in nature due to the NRI region only occurring in 
the narrow resonance band and the high gradient of the phase difference in this band. In this 
paper we seek to show how the functional bandwidth can be increased by utilisation of the 
Pancharatnam method \cite{Pancharatnam1955} that has been successfully used in the past 
with birefringent materials \cite{Pisano2006}.

\psection{Cell Geometry and Design}
Metal mesh structures can be modelled by isolating a single cell that is in practice reproduced
periodically in a two-dimensional array. The electromagnetic properties of the cell can be 
optimised using finite element analysis software, such as HFSS \cite{Ansoft}, imposing periodic 
boundary conditions around it. To create the required \ang{180} phase difference we based 
our design on a dog bone geometry (also referred to as ``I'' or ``H'' in the literature). The 
metal grid is made of a \SI{2}{\um} thick evaporated copper layer, supported by a 
polypropylene ($\epsilon = 2.2$) substrate. The HWP unit cell consists of a triplet of dog 
bones, with only the middle one embedded in the substrate. All the dimensions are given in 
the caption to Figure~\ref{fig:DBT_Unit_Cell}. The cell dimensions were optimised at 
\SI{92.5}{\GHz}, the centre of the W-band, in order to have transmissions and differential 
phase-shift along the axes respectively satisfying: $|S^{x,y}_{21}|^2 \geq 0.8$ and 
$\Delta\phi = \arg(S^x_{21}) - \arg(S^y_{21}) = \ang{-180}$.

The transmission properties of the optimised unit cell are shown in Figure~\ref{fig:DBT_S21}.
Around \SI{92.5}{\GHz} the gradient of the phase difference is very steep. The spectral 
region where the phase difference is within \SI{-180\pm3}{\degree} is extremely narrow, 
providing an operational fractional bandwidth of only \SI{0.3}{\percent} 
(\SIrange{92.8}{93.1}{\GHz}). In this region the transmissions in the $x$- and $y$-axes 
have mean values of 0.81 and 0.84 respectively. The refractive indices along the $x$- and 
$y$-axes are shown in Figure~\ref{fig:DBT_n}. These were calculated starting from the 
HFSS reflection and transmission coefficients and adopting the extraction methods 
discussed in \cite{Smith2002,Chen2004}. The refractive index, being a property of a bulk 
material, can only truly be defined for a set of regularly cascaded metamaterials. As each 
of our plates is only a single cell thick we have taken the effective thickness to be equal to 
the physical thickness to provide an indicative value of the refractive index. The true 
refractive index values would be smaller in magnitude as the effective thickness extends 
beyond the substrate surface to where the reflected radiation has a planar wave front. 
We see that a NRI band exists for $x$-polarized radiation above \SI{89.6}{\GHz} whilst in 
the $y$-axis the refractive index is constant.

\begin{figure}[b]
\centering
\includegraphics[height=3.3cm]{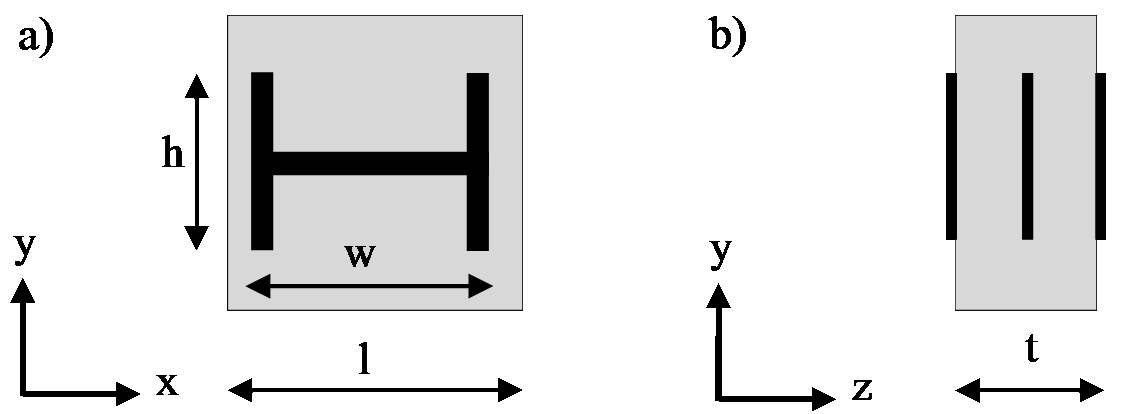}
\caption{(a)~Face on view of a unit cell. (b)~Side on view of the dog bone triplet cell. The dimensions are: $h = \SI{300}{\um}$, $w = \SI{556}{\um}$, $l = \SI{591}{\um}$ and $t = \SI{260}{\um}$. The copper is \SI{35}{\um} wide and \SI{2}{\um} thick.}
\label{fig:DBT_Unit_Cell}
\begin{minipage}[b]{0.49\linewidth}
\begin{center}
\includegraphics[width=\linewidth]{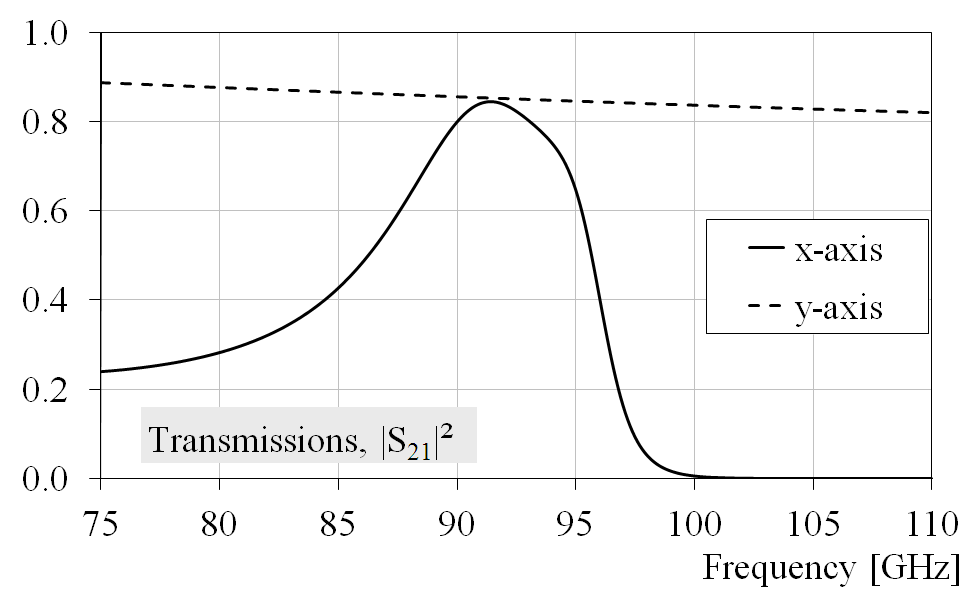}
(a)
\end{center}
\end{minipage}
\hfill
\begin{minipage}[b]{0.49\linewidth}
\begin{center}
\includegraphics[width=\linewidth]{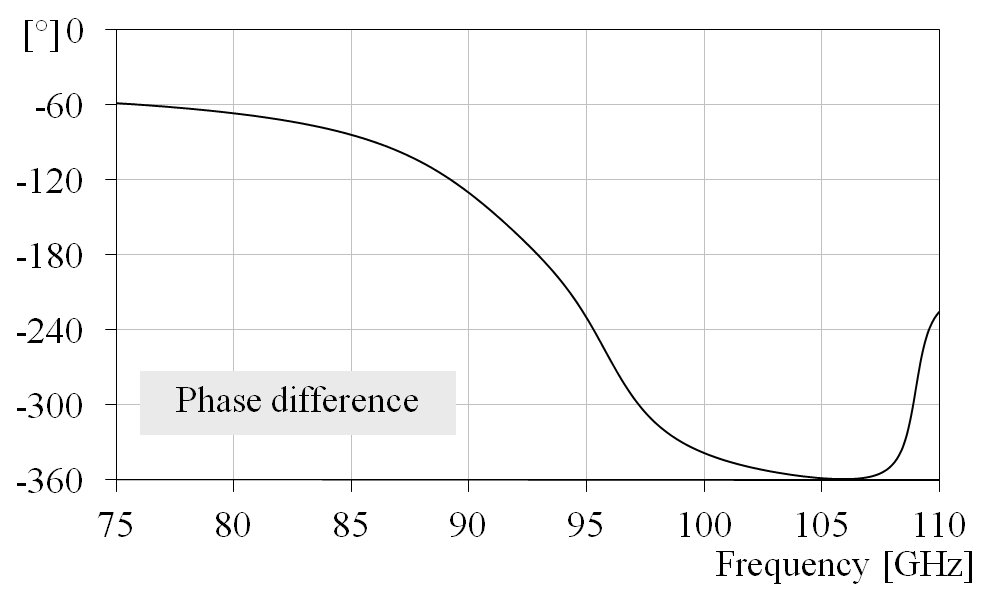}
(b)
\end{center}
\end{minipage}
\caption{(a)~Transmissions along the $x$ and $y$ axes of the one Dog Bone Triplet (DBT) cell. (b)~Phase difference between the $x$ and $y$ axes produced by a single DBT unit cell.}
\label{fig:DBT_S21}
\begin{minipage}[b]{0.49\linewidth}
\begin{center}
\includegraphics[width=\linewidth]{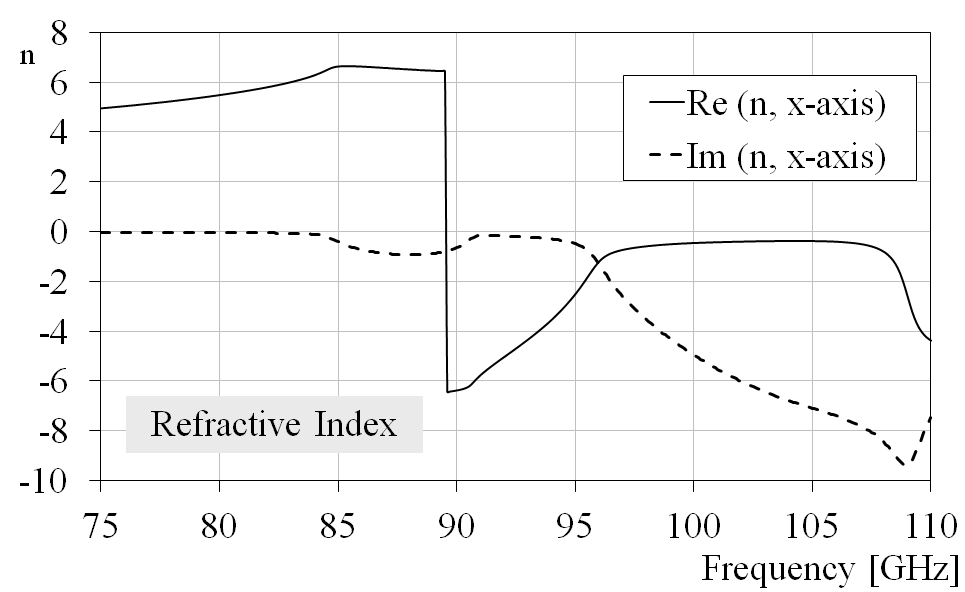}
(a)
\end{center}
\end{minipage}
\hfill
\begin{minipage}[b]{0.49\linewidth}
\begin{center}
\includegraphics[width=\linewidth]{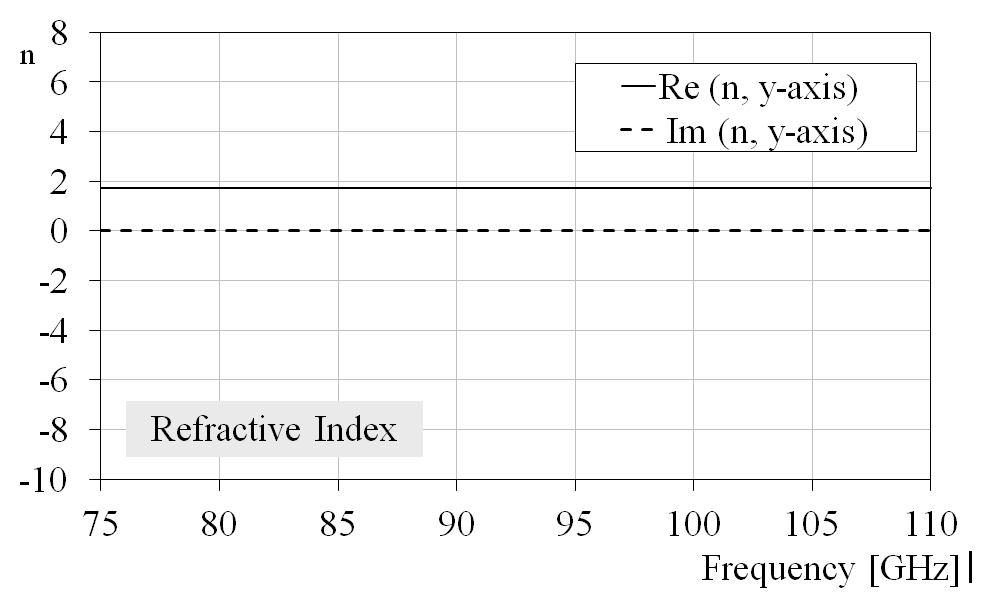}
(b)
\end{center}
\end{minipage}
\caption{(a)~Refractive index of the dog bone triplet along the $x$-axis. (b)~Refractive index along the $y$-axis..}
\label{fig:DBT_n}
\end{figure}

\psection{Modelling}
Whilst modeling of a single cell's transmission spectra in HFSS can be carried out comfortably, 
modelling of a cascaded set of cells is a very CPU and memory intensive process. In addition, 
the Pancharatnam recipes that we want to adopt are based on stacks of rotated plates. Given the 
geometry of the cell, when rotated it would not be possible to make use of the periodic boundary 
conditions provided by HFSS. In this case we turned to transmission line modeling which can 
produce accurate results provided the cells are far enough such that inter-cell interactions can 
be considered negligible. The S-parameters for a single cell are converted into the ABCD 
(Transmission) parameters \cite{Pozar1998} and the air gaps between the cells are represented 
using the propagation matrices for a transmission line. The Pancharatnam method was then 
implemented as had been done in previous papers \cite{Pisano2006} using three plates. The 
distance and angle between the successive plates was optimised to get the broadest possible 
bandwidth where the phase difference was flat and the transmission on both axes were as high and 
equal to each other as possible.

The resultant values came to be \SI{1.3}{\mm} for the air gaps and angles of
\SIlist{30;-29;30}{\degree} for the plates with respect to the final HWP equivalent fast axis. 
This increased the bandwidth from \SI{0.3}{\percent}, for the single plate, to \SI{6.6}{\percent}, 
between \SIlist{88.8;94.9}{\GHz}. This shows that despite the phase difference's initially steep 
gradient for a single cell, marked improvements can be achieved using the Pancharatnam method: 
a bandwidth more than a factor 20 wider. The transmissions in this region vary between 0.36 and 
0.72 in the $x$-axis and 0.36 and 0.73 in the $y$-axis, with the two axes showing similar 
transmission values to each other. The simulated transmission and phase differences of this HWP 
are shown as dashed lines together with experimental data in Figure~\ref{fig:HWP_S21}.

\psection{Measurements}
To create the dog bone triplets, the single dog bone grids were made using photolithographic 
techniques. The plate required three grids to be made and layered atop one another with the 
appropriate thickness of polypropylene sheets between them. Three of these plates were made and 
mounted onto aluminium rings that acted as support and provided the necessary air gaps. 

Measurements were taken using a Rhode \& Schwarz ZVA40 vector network analyser connected 
to two WR10 heads and corrugated horns to provide Gaussian shaped beams of W-band radiation. 
Eccosorb was used to cover the surfaces in the setup to reduce unwanted reflections. The 
transmission measurements along the two axes were taken individually and compared to the 
simulated data. 

\begin{figure}[b]
\begin{minipage}[b]{0.49\linewidth}
\begin{center}
\includegraphics[width=\linewidth]{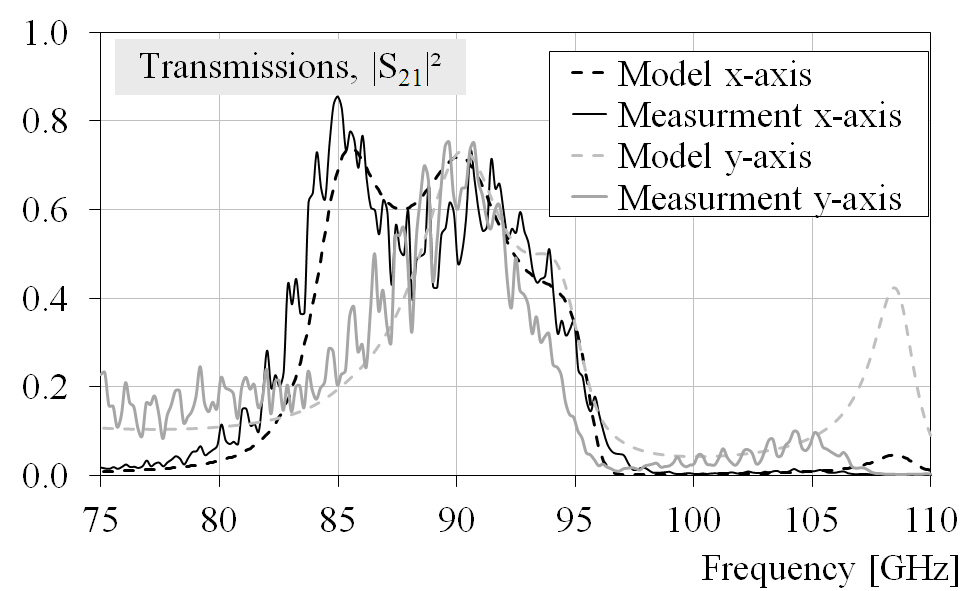}
(a)
\end{center}
\end{minipage}
\hfill
\begin{minipage}[b]{0.49\linewidth}
\begin{center}
\includegraphics[width=\linewidth]{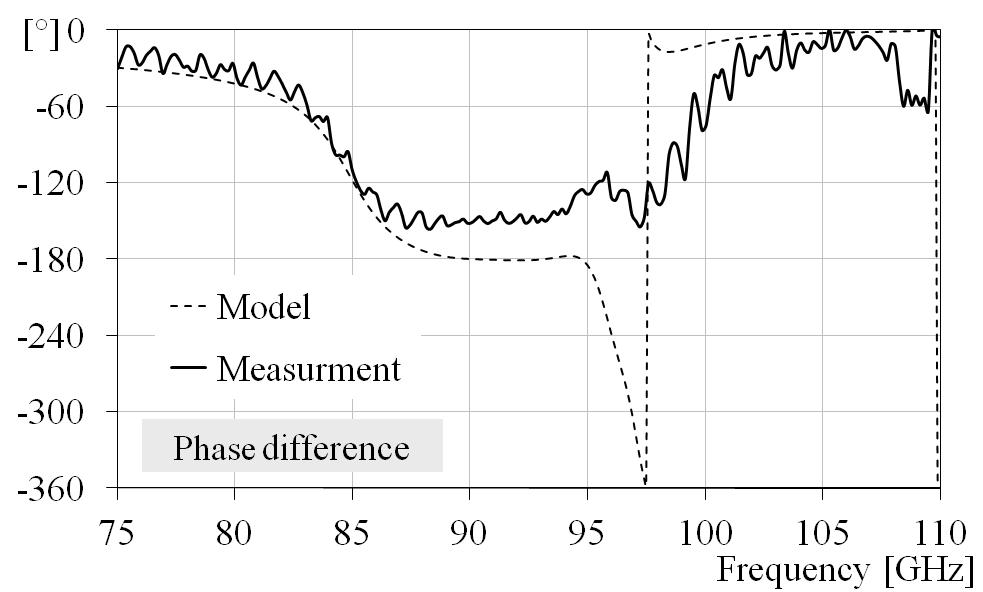}
(b)
\end{center}
\end{minipage}
\caption{(a)~The transmission on the $x$ and $y$ axes of the simulated and manufactured HWP. (b)~The phase difference.}
\label{fig:HWP_S21}
\end{figure} 

The experimentally obtained transmissions and phase differences are shown as solid lines in 
Figure~\ref{fig:HWP_S21}. Compared to the simulated transmission data, the measured 
transmission data shows a small red shift in frequency but otherwise the two show comparable 
concurrence. The phase difference also shows good correlation below \SI{87}{\GHz}. This, 
unfortunately, does not continue such that it reaches \ang{-180}. Instead, a flattening of the 
phase difference at \ang{-150} is achieved between \SI{87}{\GHz} to \SI{93.5}{\GHz} providing 
a flat region of bandwidth \SI{7.2}{\percent} in size. The difference between the simulation and 
experimental data is due to fabrication errors in the grids. Simulations have indeed showed that 
minor deviations in the air gap thicknesses and the plates' rotations would not impact on the 
HWP's ability to produce a phase difference that reached \SI{-180}{\percent}. On the other hand, 
the comparison of the model and the measured data of the individual plates showed that two of 
them suffered from deviations above \SI{85}{\GHz} in their phase difference. For example at 
\SI{93}{\GHz} where the expected phase difference is \ang{-181} these two plates had phase 
differences of \SIlist{-164;-124}{\degree}. The best performing plate achieved \ang{-176} and 
its measured phase difference is shown in Figure~\ref{fig:HWP_dp_2}. It is interesting to note 
that the average of these values equals \ang{-155}, close to the measured value of the flat 
region of the final HWP. In all of the plates, above \SI{\sim95}{\GHz}, the measured phase 
difference ceases to decrease and instead flattens out. This could explain the lack of the dip 
the measured phase difference makes above \SI{95}{\GHz} in the full HWP.

\begin{figure}[t]
\centering
\includegraphics[width=0.49\linewidth]{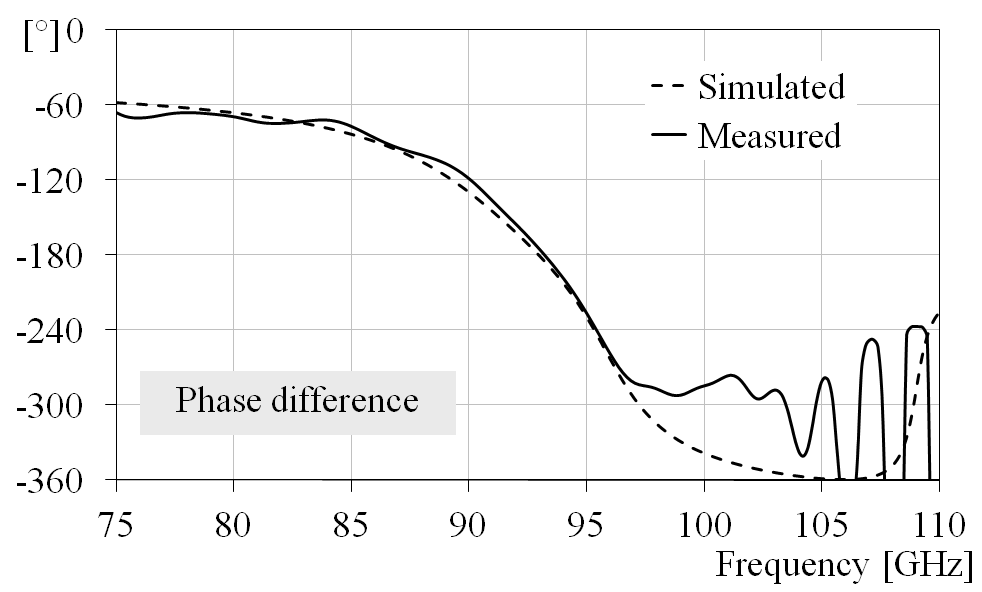}
\caption{The measured and simulated phase difference for the best performing plate.}
\label{fig:HWP_dp_2}
\end{figure}

So whilst the experimental results do not exactly match the simulations for manufacturing 
reasons, it can be seen that a broadening of a flat region in the phase difference can be 
achieved using the Pancharatnam method, even with highly birefringent wave plates. 

\psection{Conclusion}
We have created a novel HWP that utilises a highly birefringent metamaterial plate with 
refractive indices of different signs in each axis and cascaded three of them using the 
Pancharatnam method to increase their bandwidth. The simulated data showed an increase 
in bandwidth from \SI{0.3}{\percent} (\SIrange{92.8}{93.1}{\GHz}) for a single plate to 
\SI{6.6}{\percent} (\SIrange{88.8}{94.9}{\GHz}) for the whole device: about a factor 
20 wider. The experimentally measured data shows a flattening in the phase difference at 
\ang{150} between \SI{87}{\GHz} to \SI{93.5}{\GHz}. Whilst not a good match with the 
simulation this does show that in principle this method can achieve a flattening of the phase 
difference even with highly birefringent wave plates. An attempt to account for the differences 
between the simulation and experiment was made by studying the transmissions of the individual 
plates. It is found that the likely cause for the discrepancy in the final HWP's performance is 
due to fabrication errors in the individual grids themselves. Future HWP prototypes will be 
manufactured using grids selected on their performance in order to achieve the predicted 
results. Even better performances could in principle be achieved by designing single NRI plates 
with broader bandwidths, i.e.\ with geometries showing gentler phase difference slopes.

\ack
The first author is funded by a studentship from the Science and Technology Facilities Council 
(STFC).

\end{paper}

\end{document}